\documentclass[a4paper,fleqn,usenatbib]{mnras}

\usepackage[T1]{fontenc}
\usepackage{ae,aecompl}

\usepackage{graphicx}	
\usepackage{amsmath}	
\usepackage{amssymb}	

\title[Constraining the jet proper motion of J2134$-$0419]{Constraining the radio jet proper motion of the high-redshift quasar J2134$-$0419 at $z=4.3$}

\author[K. Perger et al.]{Krisztina Perger,$^{1,2}$\thanks{E-mail: k.perger@astro.elte.hu}
S\'andor Frey,$^{2}$,
Krisztina \'E. Gab\'anyi,$^{2,3}$
Tao An,$^{4,5}$\newauthor
Silke Britzen,$^{6}$
Hong-Min Cao,$^{7,8}$
D\'avid Cseh,$^{2}$
Jane Dennett-Thorpe,$^{9}$ \newauthor
Leonid I. Gurvits,$^{10,11}$
Xiao-Yu Hong,$^{4,5}$
Isobel M. Hook,$^{12}$
Zsolt Paragi,$^{10}$ \newauthor
Richard T. Schilizzi,$^{13}$
Jun Yang$^{14,4}$ 
and Yingkang Zhang$^{4}$
\\
$^{1}$Department of Astronomy, E\"{o}tv\"{o}s Lor\'{a}nd University, P\'{a}zm\'{a}ny P\'{e}ter s\'{e}t\'{a}ny 1/A, H-1117 Budapest, Hungary\\
$^{2}$Konkoly Observatory, MTA Research Centre for Astronomy and Earth Sciences, Konkoly Thege Mikl\'{o}s \'{u}t 15-17, H-1121 Budapest, Hungary\\
$^{3}$MTA-ELTE Extragalactic Astrophysics Research Group, P\'{a}zm\'{a}ny P\'{e}ter s\'{e}t\'{a}ny 1/A, H-1117 Budapest, Hungary\\
$^{4}$Shanghai Astronomical Observatory, Chinese Academy of Sciences, 80 Nandan Road, Shanghai 200030, China\\
$^{5}$Key Laboratory of Radio Astronomy, Chinese Academy of Sciences, Nanjing 210008, China\\
$^{6}$Max-Planck-Institut f\"ur Radioastronomie, Auf dem H\"ugel 69, D-53121 Bonn, Germany\\
$^{7}$School of Electronic and Electrical Engineering, Shangqiu Normal University, Wenhua Road 298, Shangqiu, Henan 476000, China\\
$^{8}$Xinjiang Astronomical Observatory, Chinese Academy of Sciences, 150 Science 1-Street, Urumqi, Xinjiang 830011, China\\
$^{9}$The University of Manchester, Manchester M13 9PL, United Kingdom\\
$^{10}$Joint Institute for VLBI ERIC, PO Box 2, NL-7990 AA Dwingeloo, The Netherlands\\
$^{11}$Department of Astrodynamics and Space Missions, Delft University of Technology, Kluyverweg 1, NL-2629 HS Delft, The Netherlands\\
$^{12}$Department of Physics, Lancaster University, Lancaster LA1 4YB, United Kingdom\\
$^{13}$Jodrell Bank Centre for Astrophysics, School of Physics and Astronomy, The University of Manchester, Manchester M13 9PL, United Kingdom\\
$^{14}$Department of Space, Earth and Environment, Chalmers University of Technology, Onsala Space Observatory, SE-43992 Onsala, Sweden\\
}

\date{Accepted 2018 XXXX YY. Received 2018 XXXX YY; in original form 2017 November 29}

\pubyear{2018}

\begin{document}
\label{firstpage}
\pagerange{\pageref{firstpage}--\pageref{lastpage}}
\maketitle

\begin{abstract}
To date, PMN J2134--0419 (at a redshift $z=4.33$) is the second most distant quasar known with a milliarcsecond-scale morphology permitting direct estimates of the jet proper motion. Based on two-epoch observations, we constrained its radio jet proper motion using the very long baseline interferometry (VLBI) technique.  The observations were conducted with the European VLBI Network (EVN) at 5~GHz on 1999 November 26 and 2015 October 6. We imaged the central 10-pc scale radio jet emission and modeled its brightness distribution. By identifying a jet component at both epochs separated by 15.86~yr, a proper motion of $\mu=0.035\pm0.023$~mas~yr$^{-1}$ is found. It corresponds to an apparent superluminal speed of $\beta_\mathrm{a}=4.1 \pm 2.7\,c$. Relativistic beaming at both epochs suggests that the jet viewing angle with respect to the line of sight is smaller than $20\degr$, with a minimum bulk Lorentz factor $\Gamma=4.3$.  The small value of the proper motion is in good agreement with the expectations from the cosmological interpretation of the redshift and the current cosmological model. Additionally we analyzed archival Very Large Array observations of J2143$-$0419 and found indication of a bent jet extending to $\sim$30~kpc.
\end{abstract}




\section{Introduction}

So far, very few multi-epoch very long baseline interferometry (VLBI) observations targeting active galactic nuclei (AGN) jets at the earliest epochs of the Universe have been made, for several reasons. First, there are only about 170 radio-emitting AGN known at $z \ga 4$ \citep[e.g.][]{perger17}, and an overwhelming majority of them are weak (mJy-level) radio sources at GHz frequencies, not suitable for detailed high-resolution high-fidelity jet studies with present-day sensitivity-limited VLBI systems. Second, the radio jets  generally have steep spectra (i.e. decreasing brightness with increasing frequency) and are less prominent at very high redshift \citep[e.g.][]{gurvits00}, simply because the emitted (rest-frame) frequency $\nu_{\rm em}$ depends on the observed frequency $\nu_{\rm obs}$ as $\nu_{\rm em} = (1+z) \nu_{\rm obs}$. Indeed, among the total of 30 radio AGN at $z \ga 4.5$ imaged with VLBI to date \citep[][ and references therein]{coppejans16}, the majority shows compact featureless structure without easily identifiable jet components on scales of milliarcseconds (mas) or tens of mas. Finally, one has to wait considerably longer for a reliable detection of a structural change in a high-redshift jet compared to a low-redshift counterpart, because the cosmological time dilation causes any jet component motion appear $(1+z)$ times slower in the observer's frame.

To date, the only direct estimates of AGN jet proper motion at $z>4$ are based on two-epoch 5-GHz VLBI observations of the blazar J1026+2542 ($z=5.3$) separated by more than 7~yr \citep[in the observer's frame,][]{frey15}. Three moving components could be identified in the jet of J1026+2542, showing proper motions $\mu \sim 0.1$~mas~yr$^{-1}$ that correspond to apparent speeds $\beta_\mathrm{a} \sim 10 c$. For another high-redshift blazar, J1430+4204 ($z=4.7$), two-epoch VLBI observations at 15 GHz could indirectly constrain $\mu$ as being no more than about 0.03~mas~yr$^{-1}$ for a putative but undetected jet component possibly emerging after a major radio outburst \citep{veres10}.   

Here we present a jet proper motion estimate for another high-redshift radio quasar, PMN~J2134$-$0419 (hereafter J2134$-$0419). This object at redshift\footnote{Another redshift measurement $z=4.346 \pm 0.005$ from \citet{hook02} is consistent within the errors.} $z=4.334 \pm 0.007$ \citep{peroux01} was observed at 5~GHz with the European VLBI Network (EVN) on 1999 November 26 and 2015 October 6.

The Green Bank Northern Sky Survey \citep{white92}, NRAO VLA Sky Survey\footnote{http://www.cv.nrao.edu/nvss/} \citep[NVSS,][]{condon98}  and Faint Images of the Radio Sky at Twenty-Centimeters\footnote{http://sundog.stsci.edu/} \citep[FIRST,][]{white97} radio surveys provide consistent radio flux density measurements of  J2134$-$0419 at 1.4~GHz of  290~mJy,  294.4$\pm$8.8~mJy and 311.3$\pm$18.3~mJy, respectively.

The latter two surveys were made with the Very Large Array (VLA) interferometer. Both the FIRST and NVSS images show an unresolved structure, although the angular resolution of FIRST ($\sim 5\arcsec$) is about an order of magnitude finer than that of the NVSS. From this limited number of 1.4-GHz flux densities, there is no clear evidence for variability. Further flux density data points are available at 365~MHz \citep[622~mJy in the Texas Survey,][]{douglas96}, at 4.85~GHz \citep[221~mJy in the Parkes--MIT--NRAO, i.e. PMN survey,][]{griffith95}, and at 20~GHz \citep[82~mJy measured with the Australia Telescope Compact Array,][]{murphy10}. 

Recently \citet{cao17} presented the results of high-resolution EVN observations of J2134$-$0419 at 1.7 and 5~GHz, including the 5-GHz data that constitute the second epoch of our proper motion study. The radio emission of the source shows a one-sided core--jet structure typical for blazars. This stucture is aligned close to the east--west direction and extends to $\sim$30~mas. The presumably self-absorbed `core' component is identified with the brighter and more compact, flat-spectrum western feature. The coordinates of J2134$-$0419 (right ascension $21^\mathrm{h} 34^\mathrm{m} 12\fs01074$ and declination $-04\degr 19\arcmin 09\farcs8610$) are accurate within 0.5~mas \citep{cao17}. The higher visibility amplitudes on the shortest baseline of the 1.7~GHz EVN data and the significantly lower flux density compared to the FIRST value led \citet{cao17} to suggest that the radio source may have additional jet structure extending to $\sim$0.1--1~arcsec angular scales. Apart from the mas-scale jet proper motion estimate based on EVN data, we also present here VLA imaging at 1.4 and 4.8~GHz using archival data obtained on 2004 October 22, to reveal the arcsec-scale radio struture of this quasar.

J2134$-$0419 is also an X-ray emitter. Using {\it Swift} X-Ray Telescope (XRT) and UltraViolet Optical Telescope (UVOT) data, \citet{sbarrato15} modeled its broad-band spectral energy distribution (SED) to obtain estimates for the the central black hole mass ($1.8 \times 10^{9}\text{ M}_{\sun}$) and the jet properties. They found solutions with $10 \la \Gamma \la 13$ for the bulk Lorentz factor and $3\degr \la \theta \la 6\degr$ for the jet viewing angle. Similar jet parameters were considered by \citet{ghisellini15} who assessed the detectability of extended lobe emission in $z>4$ blazars.

In Section \ref{sec:obs} we describe the calibration and analysis of radio interferometer (EVN and VLA) data. Section \ref{sec:results} gives the results of our imaging and brightness distribution modeling. In Section \ref{sec:disc} we derive the jet component proper motion, estimate the physical and geometric properties of the jet, and discuss the radio structure of J2134$-$0419. Summary and concluding remarks are presented in Section \ref{sec:sum}.

In this paper, we adopt the cosmological model parameters $H_0=70$~km~s$^{-1}$~Mpc$^{-1}$, $\Omega_\mathrm{M}=0.3$, and $\Omega_\Lambda=0.7$. At the redshift of J2134$-$0419, 1~mas angular size corresponds to 6.72~pc projected linear size, the luminosity distance of the quasar is 39.4~Gpc, and 1~mas~yr$^{-1}$ proper motion corresponds to 116.8 $c$ apparent transverse speed \citep{wright06}.

\section{Observations and data reduction}
\label{sec:obs}

\subsection{European VLBI Network}

The source J2134$-$0419 was first observed at 5~GHz with VLBI using the EVN among a sample of high-redshift quasars on 1999 November 26 (project code ES034B, PI: I.A.G. Snellen). The participating 9 radio telescopes were Effelsberg (Germany), Jodrell Bank Mk2 (United Kingdom), Medicina (Italy), Noto (Italy), Toru\'{n} (Poland), Onsala (Sweden), Hartebeesthoek (South Africa), Sheshan (China), and the phased array of the Westerbork Synthesis Radio Telescope (the Netherlands) with 13 antenna elements. The observing time spent on J2134$-$0419 was 3.3~h. With a recording data rate of 256 Mbit s$^{-1}$, 4 basebands (or intermediate frequency channels, IFs) were used in both left and right circular polarization, each with 16 spectral channels. The total bandwidth was 32 MHz in both polarizations. The correlation took place at the EVN Data Processor at the Joint Institute for VLBI ERIC (JIVE) in Dwingeloo (the Netherlands). The correlator integration time was 4~s.

The second observing epoch was 2015 October 6 \citep[project code EC054B, PI: H.-M. Cao,][]{cao17}. The EVN telescopes in this 5-GHz experiment were the same as above with the exception that Effelsberg did not participate, and only a single dish participated in Westerbork. The on-source time for J2134$-$0419 was 2.5~h. These observations were done in e-VLBI mode, with 1024 Mbit~s$^{-1}$ data rate, 8 IFs per polarization.
The total bandwidth was 128 MHz per polarization. The identical observing frequency band and the similar radio telescope array used at the two different epochs nearly 16 years apart result in similar restoring beam sizes and shapes and thus provide an ideal opportunity for comparing the source structures.

The data calibration and analysis of the second-epoch measurements have been done and described in detail by \citet{cao17}. The data from the 1999 experiment were  calibrated by us in the US National Radio Astronomy Observatory (NRAO) Astronomical Image Processing System\footnote{http://www.aips.nrao.edu/index.shtml} 
\citep[{\sc aips},][]{greisen2003} in a standard way \citep[e.g.][]{diamond1995}. A priori 
amplitude calibration was done using the antenna gain curves and the system temperatures supplied by 
the EVN stations as measured during the observations. Manual phase calibration using a short scan on a bright calibrator source and then global fringe-fitting \citep{schwab1983} for J2134$-$0419 were performed. The calibrated visibility data were exported for imaging in the {\sc difmap} program \citep{shepherd97}. The data were averaged over 30-s intervals and inspected for outliers that were flagged in a baseline-based manner. The data from Toru\'{n} had to be discarded because of amplitude calibration problems. Cycles of conventional hybrid mapping were performed in {\sc difmap} with {\sc clean} component modeling \citep{hogbom1974} and phase (later also amplitude) self-calibration. Finally, Gaussian brightness distribution model components were fitted to the self-calibrated visibility data in {\sc difmap}, to obtain estimates for component sizes, positions and flux densities. This was done for the 2015 data set as well, allowing for a quantitative comparison of the models to look for any noticeable change between the two epochs.

\subsection{Very Large Array}

The VLA observations of J2134$-$0419 were conducted at 1.4 and 4.8~GHz frequencies on 2004 October 22 (project code: AC755C). We obtained the data from the NRAO Science Data Archive\footnote{http://archive.nrao.edu} and loaded them into {\sc aips} for calibration. The VLA was used in its most extended A configuration providing the highest angular resolution. The on-source integration times were nearly 38~min and 12~min at 1.4~GHz and 4.8~GHz, respectively, and the total bandwidth was 100~MHz. The phases and amplitudes were calibrated in {\sc aips} in a standard way as prescribed in the Cookbook\footnote{http://www.aips.nrao.edu/cook.html}, using 3C\,48 as the primary flux density calibrator. Imaging and model-fitting were performed in {\sc difmap}.

\section{Results}
\label{sec:results}

\subsection{EVN data}

\begin{figure}
\centering
\includegraphics[width=\linewidth]{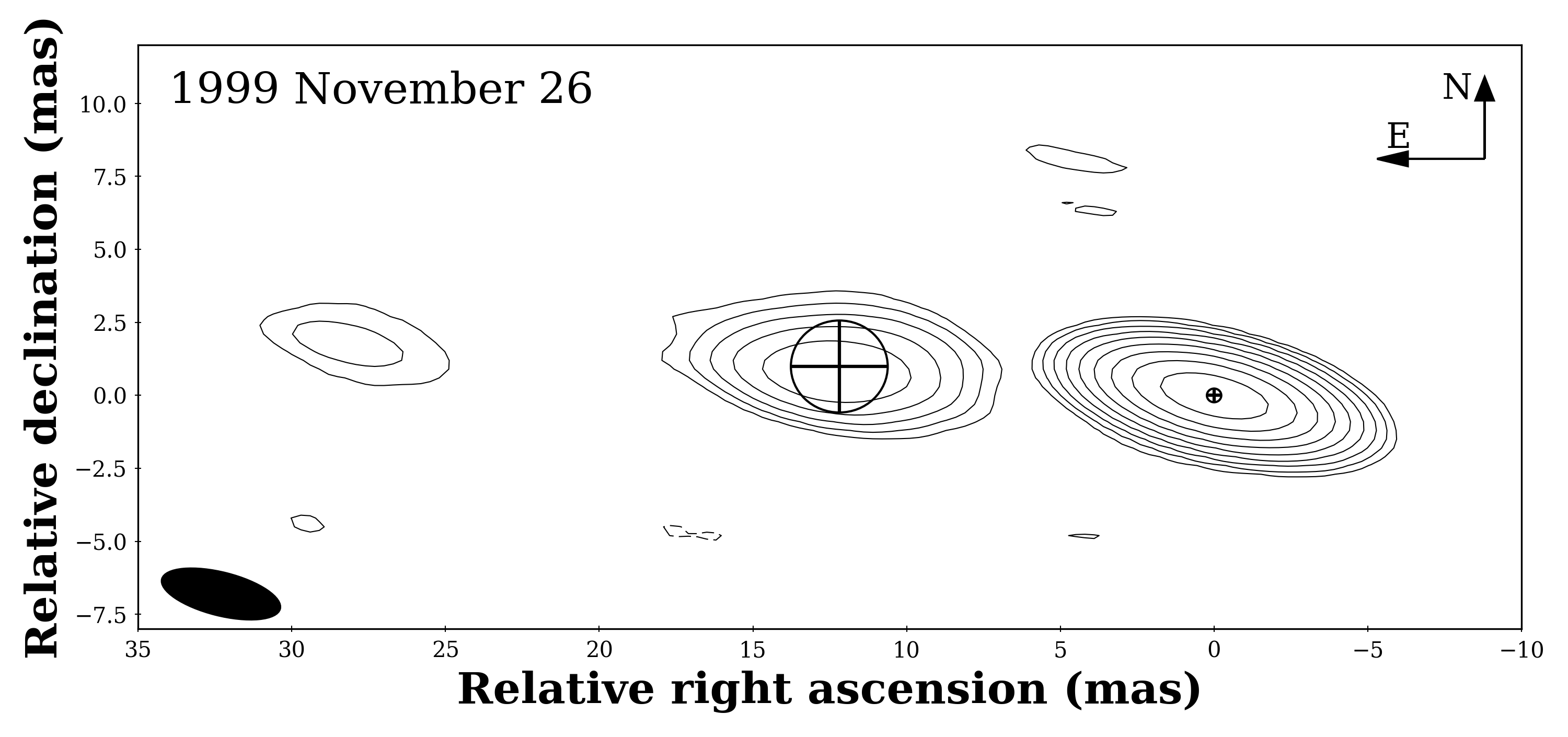}
\vspace*{5pt}

\includegraphics[width=\linewidth]{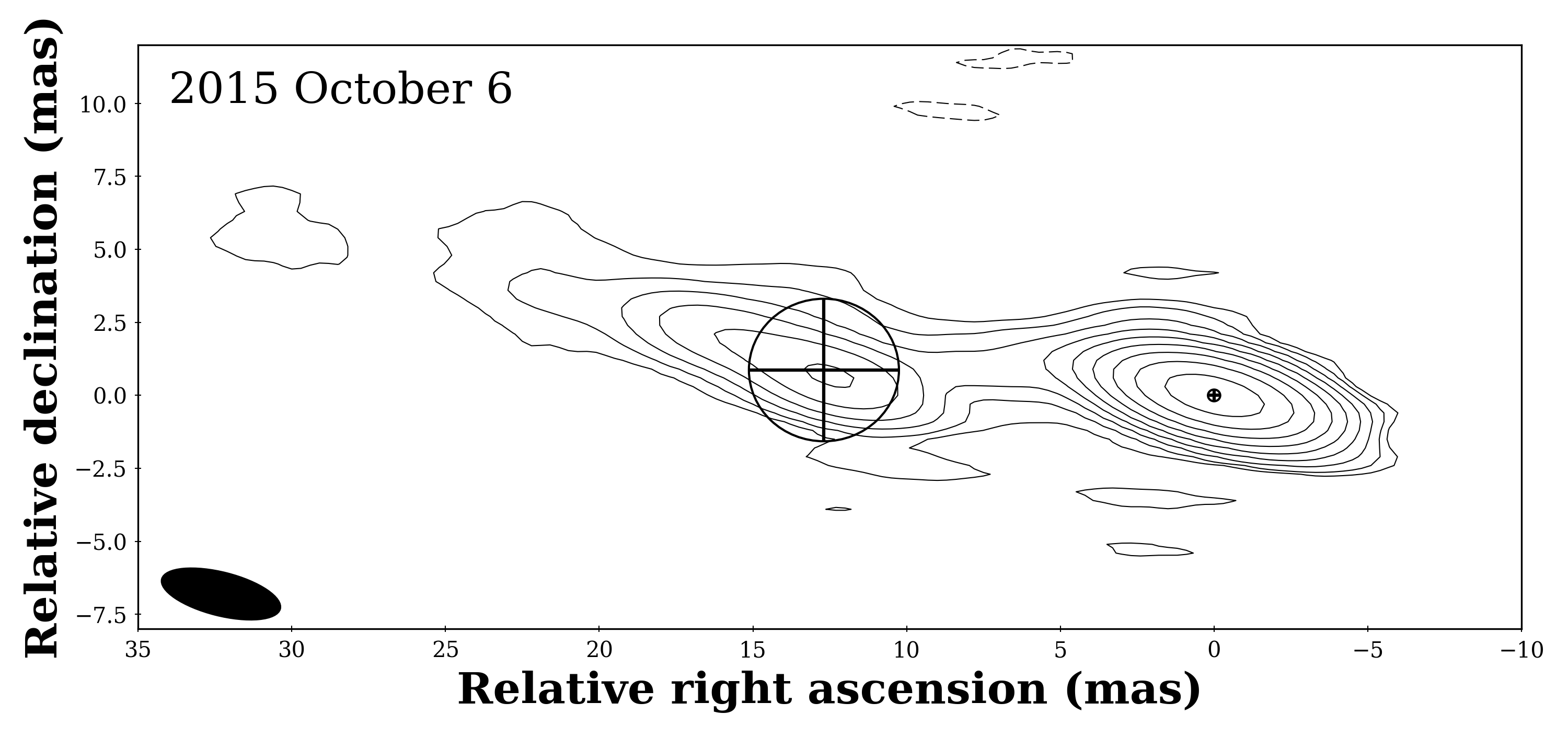}
\caption{Naturally weighted 5-GHz {\sc clean} images of the quasar J2134$-$0419. The 1999 and 2015 EVN images are shown in the upper and lower panels, respectively. The elliptical Gaussian restoring beam size in both cases was set to $1.5\,\mathrm{mas} \times 4.0\,\mathrm{mas}$ (FWHM) at $75\degr$ major axis position angle (measured from north through east), as shown in the bottom left corner of the panels. The lowest contour levels are drawn at $\pm 0.26$~mJy~beam$^{-1}$ and $\pm 0.38$~mJy~beam$^{-1}$ for the first and second epochs, respectively, corresponding to $\sim$3$\sigma$ image noise. Further contours increase by a factor of 2. The peak intensity is 115.3~mJy~beam$^{-1}$ at the 1999 epoch and 144.3~mJy~beam$^{-1}$ at the 2015 epoch. The crossed circles indicate the sizes and positions of the fitted Gaussian components.}
\label{fig:EVN}
\end{figure}

The naturally weighted EVN images of J2134$-$0419 at both epochs are shown in Fig.~\ref{fig:EVN}. The relative coordinates are centered at the brightness peak. For an easy comparison of the two images, the same restoring beam was used in both cases. Since the radio telescope arrays and the observing times were similar, and the sidereal time ranges were overlapping, the $(u,v)$ coverages did not differ substantially at the two epochs in 1999 and 2015. Note that the image restored with a smaller beam from the same data from 2015 \citep[Fig.~2b of][]{cao17} is very similar, both qualitatively and quantitatively. The images in Fig.~\ref{fig:EVN} clearly show the known compact radio structure with two prominent features, the westernmost `core' and a jet component to its east. At both epochs, we fitted the observed brightness distributions by two Gaussian components using the visibility data \citep[e.g. ][]{pearson95}, to describe the `core' and the jet components in a simple way. There is a hint on a third, outermost component at about 30 mas from the `core' in Fig.~\ref{fig:EVN}. We also attempted to model it but the fit to the visibility data did not provide a useful solution, most likely because the underlying feature is too faint and diffuse. The reduced $\chi^2$ values indicating the goodness of fit did not decrease significantly when adding a third model component.

The model parameters allow us to characterise the changes in the position of the jet component between 1999 and 2015. The  parameters of the fitted model components including their relative positions are listed in Table~\ref{tab:VLBI}. We estimated the uncertainties of the fitted model parameters following \citet{lee08}, considering an additional 5 per cent error in quadrature for the flux densities to account for the uncertainties of the VLBI amplitude calibration. Note that the values in Table~\ref{tab:VLBI} for the second epoch in 2015 agree well within the errors with those determined by \citet{cao17} using the same EVN data.  

\begin{table*}
\centering
\caption{VLBI image and model fitting parameters of J2134$-$0419 at 5~GHz.}
\label{tab:VLBI}
\begin{tabular}{ccccccc}
\hline\hline
Epoch & Component & $P$ (mJy~beam$^{-1}$) & $S$ (mJy)& $\vartheta$ (mas) & $R$ (mas)&$\phi$ ($^\circ$)\\
\hline
1999 & core &115.3$\pm$3.8& 121.2$\pm$5.5  & 0.47$\pm$0.02 & 0 & 0\\
     & jet  &  9.0$\pm$1.1&  14.9$\pm$2.2  & 3.15$\pm$0.38 & 12.19$\pm$0.19 & 85.3$\pm$0.9\\
2015 & core &144.3$\pm$5.3& 159.2$\pm$11.2 & 0.41$\pm$0.02 & 0 & 0\\
     & jet  &  9.8$\pm$1.4&  26.3$\pm$4.2  & 4.59$\pm$0.65 & 12.72$\pm$0.32 & 86.1$\pm$1.5\\
\hline
\end{tabular}
\\
{\it Notes.} Column 3 -- image peak intensity, Column 4 -- model component flux density, Column 5 -- circular Gaussian model component size (full width at half-maximum, FWHM), Column 6 -- angular radial distance from the core, Column 7 -- position angle with respect to the core, measured from north through east.
\end{table*}

\subsection{VLA data}
\begin{figure}
\centering
\includegraphics[width=0.8\linewidth]{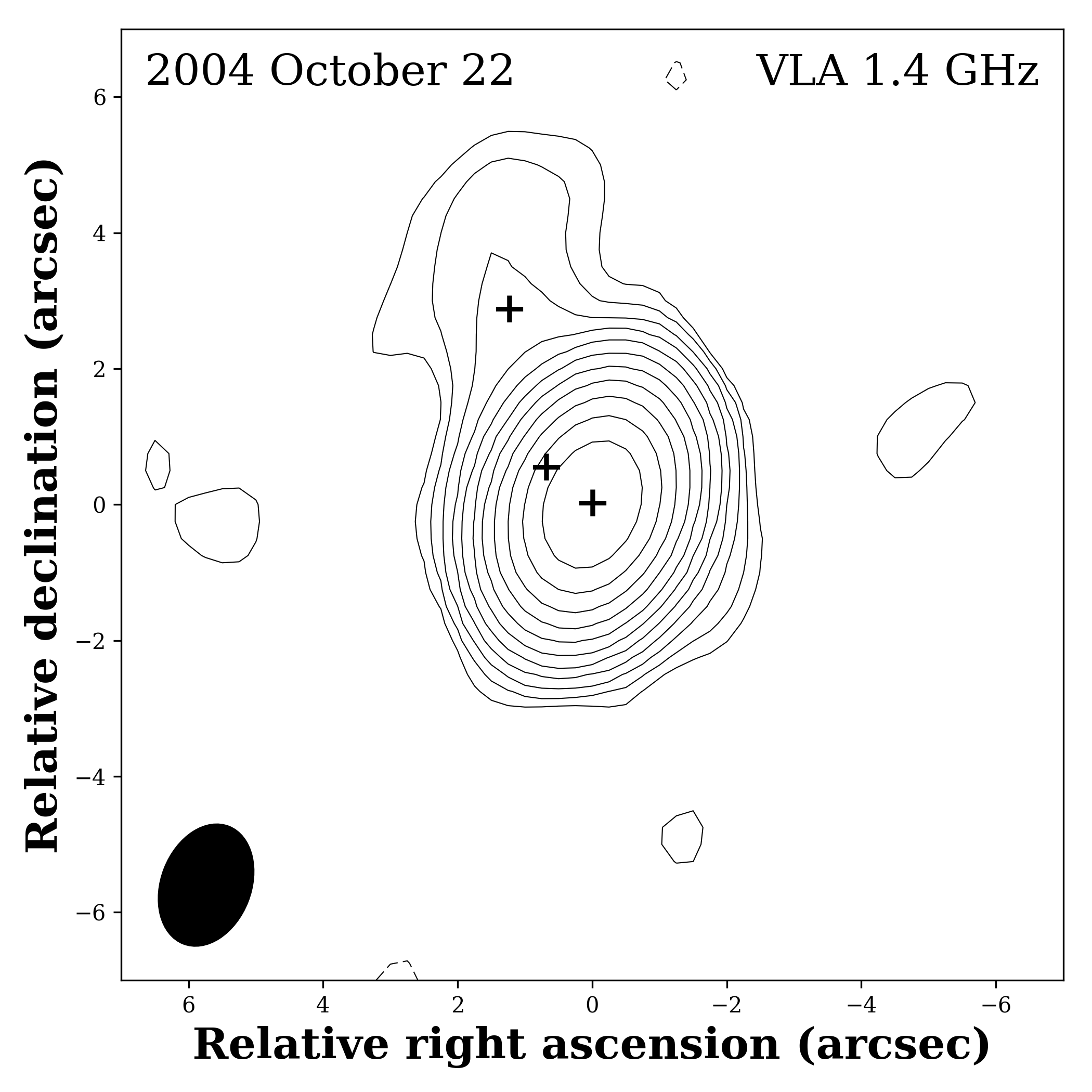}
\caption{Naturally weighted 1.4-GHz VLA image of J2134$-$0419. The lowest contour levels are drawn at $\pm0.14$~mJy~beam$^{-1}$, corresponding to $\sim$3$\sigma$ image noise. Positive contours increase by a factor of 2.  The peak brightness is 309.2~mJy~beam$^{-1}$. The restoring beam size is $1\farcs3\times 1\farcs9$ at $-22\degr$ position angle. The fitted model components are indicated with crosses (see Table~\ref{tab:VLA}).}
\label{fig:VLA-L}
\end{figure}

\begin{figure}
\centering
\includegraphics[width=0.8\linewidth]{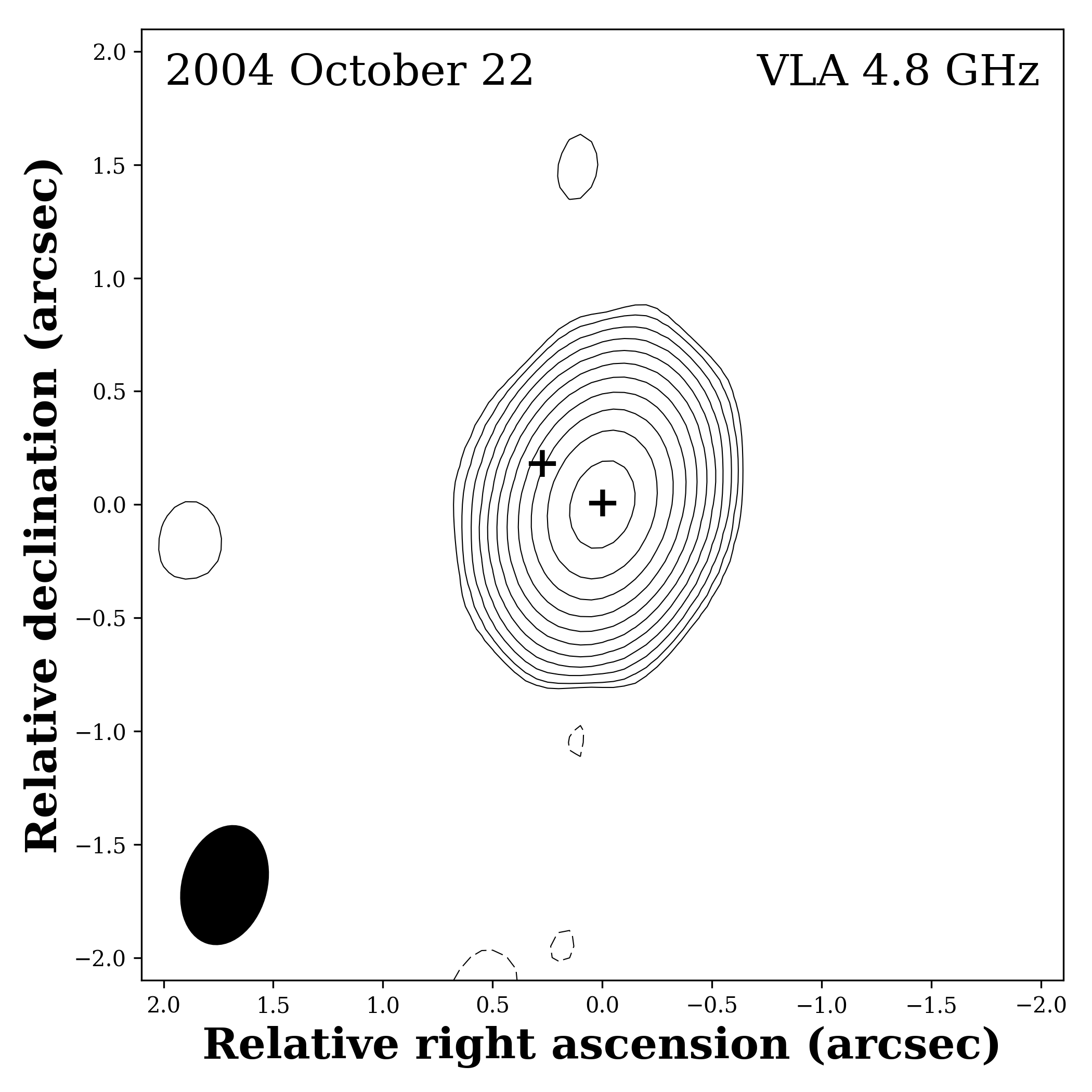}
\caption{Naturally weighted 4.8-GHz VLA image of J2134$-$0419. The lowest contour levels are drawn at $\pm0.15$~mJy~beam$^{-1}$, corresponding to $\sim$3$\sigma$ image noise. Positive contours increase by a factor of 2. The peak brightness is 224.1~mJy~beam$^{-1}$. The restoring beam size is $0\farcs4\times 0\farcs5$ at $-17\degr$ position angle. The fitted model components are indicated with crosses (see Table~\ref{tab:VLA}).}
\label{fig:VLA-C}
\end{figure}

The naturally weighted VLA images at 1.4~GHz and 4.8~GHz are shown in Fig.~\ref{fig:VLA-L} and Fig.~\ref{fig:VLA-C}, respectively. The 1.4-GHz image indicates a weak jet-like feature to the north-northeast of the core, extending up to $\sim 5$~arcsec. The 4.8-GHz image shows a compact, practically unresolved source. We fitted circular Gaussian model components to the VLA visibility data. The model parameters are listed in Table~\ref{tab:VLA}. In Figs.~\ref{fig:VLA-L} and \ref{fig:VLA-C}, we also marked the positions of these components.

\begin{table*}
\centering
\caption{VLA  image and model fitting parameters of J2134$-$0419 measured in 2004 October.}
\label{tab:VLA}
\begin{tabular}{ccccccc}
\hline\hline
$\nu_\mathrm{obs}$ (GHz) & Component & $P$ (mJy~beam$^{-1}$) & $S$ (mJy)& $\vartheta$ (mas) & $R$ (mas)&$\phi$ ($^\circ$)\\
\hline
1.4 & 1	& 309.2$\pm$6.8	& 308.7$\pm$9.6	& $<20$		& 0		&0 \\
	& 2	& 1.6$\pm$0.5	& 1.6$\pm$0.7	& $<270$	& 864$\pm$31	&52.2$\pm$2.0\\
	& 3 & 0.7$\pm$0.4	& 1.8$\pm$1.0	& 2035$\pm$1065	& 2856$\pm$533	&23.4$\pm$9.7\\
4.8	& 1	& 224.1$\pm$6.2	& 225.8$\pm$8.8	& 39.2$\pm$1.1		& 0		& 0	\\	
	& 2	& 0.7$\pm$0.4	& 0.6$\pm$0.5	& $<130$		& 326.6$\pm$3.0		&57.8$\pm$0.5\\
\hline
\end{tabular}
\\
{\it Notes.} Column 1 -- observing frequency, Column 3 -- image peak intensity (measured in the residual image for the jet components), Column 4 -- model component flux density, Column 5 -- circular Gaussian model component size (FWHM) or upper limit corresponding to the minimum resolvable angular size \citep{kovalev05}, Column 6 -- angular radial distance from the core, Column 7 -- position angle with respect to the core, measured from north through east.
\end{table*}

\section{Discussion}
\label{sec:disc}

\subsection{Radio structure}

At both EVN epochs, an indication of the continuation of the jet up to $\sim$30~mas (corresponding to a projected distance of 84~pc)  from the core is apparent (Fig.~\ref{fig:EVN}), albeit at low brightness levels. While the two major features are distinct in 1999, a more continuous jet flow is visible in the 2015 image. Also, modelfit results (Table~\ref{tab:VLBI}) suggest that the jet component became slightly more extended from 1999 to 2015. 

On larger scales, from our VLA A-configuration images at 1.4~GHz (Fig.~\ref{fig:VLA-L})  and at 4.8~GHz (Fig.~\ref{fig:VLA-C})  and from the position angles of the corresponding model components (Table~\ref{tab:VLA}), it seems that the jet continues out to about 5 arcsec, but its position angle gradually turns northward. Such a misalignment between the inner and outer (arcsec-scale) jet often attributed to helical jet geometry is not uncommon for blazars \citep[e.g.][]{conway93,kharb10,zhao11,singal16}. 

The different flux densities measured with the EVN at the two epochs (Table~\ref{tab:VLBI}) clearly indicate variability  by 30 per cent. The sum of the flux densities in the VLBI components (136.1~mJy in 1999 and 185.5~mJy in 2015) are lower than the total flux density from either the PMN survey \citep[221~mJy,][]{griffith95} or our VLA measurement ($224.8 \pm 6.2$~mJy, Table~\ref{tab:VLA}) in the 5-GHz band. This difference could be attributed to variability, although more likely it is due to jet structure on scales <300~mas that is not resolved by the VLA but is completely resolved by the VLBI baselines. The flux densities we measured with the VLA (Table~\ref{tab:VLA}) are identical within the errors with the FIRST value at 1.4~GHz, and the PMN survey measurement at 4.8~GHz.

\subsection{Jet proper motion}

Based on the relative positions of the jet component with respect to the `core' deduced from our fitted brightness distribution models (Table~\ref{tab:VLBI}), we estimated the proper motion of the jet component over the 5793~d (15.86~yr) between the two epochs. The $0.56\pm0.37$~mas change in the angular distance translates to $\mu=0.035\pm0.023$~mas~yr$^{-1}$ proper motion. This estimate is based on two epochs only and therefore should be treated with caution, but even such a tentative value for a quasar at $z>4$ is almost unique \citep{frey15}. The jet component proper motion in J2134$-$0419 corresponds to a superluminal apparent speed $\beta_\mathrm{a}=4.1\pm2.7\,c$. The relatively slow proper motion is in good agreement with the values expected for the highest-redshift sources, dictated by the current cosmological models \citep[e.g.][]{vermeulen94,kellermann99}, and the proper motions that can be extrapolated from VLBI measurements of lower-redshift AGN samples \citep[e.g.][]{kellermann04,britzen08, hovatta08, lister09}. It is also comparable to the moderate proper motion ($\sim$0.1~mas~yr$^{-1}$) recently found in the $z=5.3$ blazar J1026+2542 \citep{frey15}. The value determined for J2134$-$0419 is well under the envelope on the $\mu-$z diagram shown in Fig. 5. of \cite{frey15}.

\subsection{Lorentz factor and inclination angle}

To see how much the innermost jet region of J2134$-$0419 is affected by relativistic beaming, we determined the `core' brightness temperatures at both epochs from the fitted model parameters (Table~\ref{tab:VLBI}) using the following expression \citep[e.g.][]{condon82}:
\begin{equation}
T_\mathrm{b}=1.22 \times 10^{12} \, (1+z) \, \frac{S}{\vartheta^2\nu^2}\,\text{K},
\end{equation}
where the flux density $S$ is measured in Jy, the circular Gaussian component angular diameter (full width at half-maximum, FWHM) $\vartheta$  in mas, and the observing frequency $\nu$ in GHz. The  measured brightness temperatures are $T_{\mathrm b,1999} = (1.5\pm0.2) \times 10^{11}$~K and $T_{\mathrm b,2015} = (2.5\pm0.4) \times 10^{11}$~K for the 1999 and 2015 epochs, respectively. From these, we could estimate the Doppler factors using

\begin{equation} 
\delta=\frac{T_\mathrm{b}}{T_\mathrm{b,int}},
\end{equation}
where $T_\mathrm{b,int}$ denotes the intrinsic brightness temparature. We obtained lower and upper limits to the Doppler factors, by adopting the value $T_\mathrm{b,int} \approx 5\times 10^{10}$~K at energy equipartition between the radiating particles and the magnetic field \citep{readhead94} on the one hand, and by assuming $T_\mathrm{b,int} \approx 3 \times 10^{10}$~K as determined by \citet{homan06} for a sample of AGN jets in a median-low brightness state on the other hand. The corresponding Doppler factors are $3 \la \delta_{1999} \la 5$ and $5 \la \delta_{2015} \la 8.3$. 

Parameters of the jet, Lorentz factor, $\Gamma$, and viewing angle to the line of sight, $\theta$, were determined from the formulae

\begin{equation}
\Gamma=\frac{\beta_\mathrm{a}^2+\delta^2+1}{2\delta}
\end{equation}

\begin{equation}
\tan\theta=\frac{2\beta_\mathrm{a}}{\beta_\mathrm{a}^2+\delta^2-1}
\end{equation}

\noindent given e.g. by \citet{urry95}.

The values for Lorentz factors allowed by both the Doppler factors and by the apparent speed (with errors) span the range $2 \la \Gamma \la 7$, while the viewing angles span the approximate range $3\degr \la \theta \la 20\degr$. Using $\beta_\mathrm{a}=4.1\,c$, the bulk Lorentz factor and viewing angle are $4.3 \le \Gamma_{1999} \le 4.5$ and $11\fdg4\le \theta_{1999} \le 18\fdg3$, and $4.3 \le \Gamma_{2015} \le 5.2$ and $5\fdg5 \le \theta_{2015} \le 11\fdg4$ for the 1999 and 2015 epochs, respectively.

Our estimates for the bulk Lorentz factor based on high-resolution radio interferometric observations are somewhat lower than the $10 \le \Gamma \le 13$ values found from SED modeling by \citet{sbarrato15}. This might be caused by the uncertainties inherent to the two independent methods, time variability, the different jet regions probed by VLBI and X-ray measurements  (since X-ray emission can be produced in many places, e.g. the corona, the inner jet or the kpc-scale jet and also the lobe emission may have an important contribution), or if the innermost VLBI `core' component is blended with an unresolved jet component, causing its fitted FWHM size appear somewhat larger \citep[see][]{natarajan17}.   In any case, our values lie close to the typical Lorentz factors $5 \la \Gamma \la 15$ applied by \citet{volonteri11} for $z>3$ radio-loud quasars in general. 

\section{Summary and conclusions}
\label{sec:sum}
We analysed VLBI imaging observations  on 1999 November 26 and 2015 October 6 \citep{cao17}, at two epochs separated by 15.86~yr, with the aim of constraining the jet proper motion of the high-redshift quasar J2134$-$0419. 
Taking the cosmological time dilation into account, this period is equivalent to 2.97~yr in the rest frame of the source. By modeling the intensity distribution of the mas to 10-mas scale compact radio structure of the source with circular Gaussian components (Table~\ref{tab:VLBI}), we tentatively determined the proper motion of a jet component as $\mu=0.035\pm0.023$~mas~yr$^{-1}$. This corresponds to a superluminal apparent speed $\beta_\mathrm{a}=4.1\pm 2.7\,c$. The jet component proper motion is slow compared to typical values measured in low-redshift blazars, in accordance with the expectations from cosmological considerations, and is consistent with the limited data available for other high-redshift AGN, as it lies well under the envelope in the $\mu-z$ diagram \citep{frey15}. 

We also calculated the Doppler-boosting factor from the measured brightness temperatures and estimated the Lorentz factor of the bulk jet flow, as well as the jet inclination angle to the line of sight for J2134$-$0419. The ranges of these quantities are $2 \la \Gamma \la 7$ and $3\degr \la \theta \la 20\degr$. (However, these are not independent ranges.) The VLBI-derived Lorentz factor is somewhat smaller and the viewing angle is larger than obtained by SED modeling \citep{sbarrato15}, but both methods support the blazar classification of J2134$-$0419. Better constraints on the proper motion and the jet properties could be obtained with repeated VLBI observations of this source with time differences of several years. Considering the two epochs of the EVN observations, J2134$-$0419 shows clear flux density variability in the compact radio structure between 1999 and 2015, as expected for blazars.

The arcsec-scale radio structure of J2134$-$0419 obtained from unpublished VLA A-configuration data taken at 1.4 and 4.8~GHz shows that the structure is dominated by
a compact feature, with a weak extension up to $\sim$5~arcsec to the north-northeast as seen in the 1.4-GHz VLA image (Fig.~\ref{fig:VLA-L}). From model fitting, there is clear indication of a gradual bending of the jet, with a misalignment of more than $60\degr$ between mas to arcsec scales. The difference between the integrated 5-GHz radio flux density of the components detected with the EVN and VLA suggests the presence of a jet on scales of $\sim100$ to 300~mas. This could be verified with intermediate-resolution interferometric imaging using the e-MERLIN and SKA1-mid arrays in the future.  

For a comprehensive statistical study of high-redshift quasar jet proper motions and jet properties, a much larger sample would be needed. Including J2134$-$0419 reported here, there are only two quasars at $z>4$ known to date whose jet proper motions have been estimated using two-epoch VLBI imaging observations \citep{frey15}. Investigating the redshift dependence of jet proper motions of an ensemble of high-redshift ($z>3-4$) sources would potentially place useful constraints on the cosmological model parameters \citep{kellermann99}. 
On the other hand, knowledge of the parameters of jets in an extensive sample of high-redshift quasars would offer invaluable clues to tackle the problem of the apparent deficit of unbeamed jetted quasars in the early Universe \citep{volonteri11,ghisellini15,ghisellini16}. For reliable detections of the slow apparent changes in their structure, high-redshift jets typically require long VLBI observational history. After several decades of VLBI measurements and surveys, studies of kinematic and physical properties of very distant quasar jets are now becoming feasible lately, helped by technical developments that allow increased sensitivity to the relatively faint emissions at high redshift. 

\section*{Acknowledgements}

The EVN is a joint facility of independent European, African, Asian and North American radio astronomy institutes. Scientific results from data presented in this publication are derived from the following EVN project codes: ES034 and EC054. The authors acknowledge the efforts and cooperation of the PI of the EVN project ES034, Ignas Snellen, and other co-investigators.
The NRAO is a facility of the National Science Foundation operated under cooperative agreement by Associated Universities, Inc.
KP, SF, K\'{E}G and DC thank the Hungarian National Research, Development and Innovation Office (OTKA NN110333) for support. This work was supported by the China--Hungary Collaboration and Exchange Programme by the International Cooperation Bureau of the Chinese Academy of Sciences.
K\'{E}G was supported by the J\'{a}nos Bolyai Research Scholarship of the Hungarian Academy of Sciences. HMC acknowledges support by the program of the Light in China's Western Region (Grant No. 2016-QNXZ-B-21) and the National Science Foundation of China (Grant No. U1731103 and 11573016).






\bsp	
\label{lastpage}
\end{document}